\documentclass[usenatbib]{mnras}

\usepackage{graphicx}	
\usepackage{amsmath}	
\usepackage{amssymb}	

\def\upjs{Institute of Physics, Faculty of Science, Pavol Jozef \v{S}af\'{a}rik University, 040 01 Ko\v{s}ice, Slovakia}
\def\tatry{Astronomical Institute, Slovak Academy of Sciences, 059 60 Tatransk\'a Lomnica, Slovakia}

\title{Chaos in multiplanetary extrasolar systems}

\author[P. Gajdo\v{s} \& M. Va\v{n}ko]{
Pavol Gajdo\v{s},$^{1}$\thanks{E-mail: pavol\,.\,gajdos\,@\,upjs\,.\,sk}
Martin Va\v{n}ko$^{2}$
\\
$^{1}$\upjs\\
$^{2}$\tatry
}

\date{Accepted 2022 November 02. Received 2022 November 02; in original form 2022 June 27.}

\pubyear{2023}

\begin{document}
\setcounter{page}{2068}
\label{firstpage}
\volume{518}
\pagerange{\pageref{firstpage}--\pageref{lastpage}}
\maketitle

\begin{abstract}
Here we present an initial look at the dynamics and stability of 178 multiplanetary systems which are already confirmed and listed in the NASA Exoplanet Archive. To distinguish between the chaotic and regular nature of a system, the value of the MEGNO indicator for each system was determined. Almost three-quarters of them could be labelled as long-term stable. Only 45 studied systems show chaotic behaviour. We consequently investigated the effects of the number of planets and their parameters on the system stability. A comparison of results obtained using the MEGNO indicator and machine-learning algorithm SPOCK suggests that the SPOCK could be used as an effective tool for reviewing the stability of multiplanetary systems. A similar study was already published by Laskar and Petit in 2017. We compared their analysis based on the AMD criterion with our results. The possible discrepancies are discussed. 
\end{abstract}


\section{Introduction} 
In recent years, we have seen an increasing number of planetary systems where additional planets have been discovered. In order to understand their architecture, we have to study planetary systems in terms of stability and dynamics. During the last two decades, several studies have been devoted to the dynamics of exoplanetary systems using various dynamical models and numerical techniques \citep[e.g.][]{Lee2004,Ferraz2006,Hadjidemetriou2006,Henrard2008,Voyatzis2008,Vaneylen2019,Gilbert2020}. Distinguishing between chaotic and regular systems is one of the main tasks in the study of dynamical systems in general. The key point is to identify areas of system parameters where instabilities can occur. One important aspect of the chaotic system behaviour is its strong dependence on the initial conditions of the numerical integration. In celestial mechanics, this input includes the orbital elements of all bodies in the system for a given epoch. Then even a small change leads to diametrically different system evolution. This fact uses several indicators of chaos e.g. Lyapunov Characteristic Exponents \citep{Cincotta2003,Skokos2010,Rein2016} or Mean Exponential Growth factor of Nearby Orbits \citep[MEGNO;][]{Cincotta2000,Godziewski2001,Cincotta2003,Marzari2021}. Not all indicators are based on this initial-condition dependence. Reversibility Error Method \citep{Panichi2014,Panichi2017}, Angular Momentum Deficit \citep[AMD;][]{Laskar2017,Petit2017,He2020} or Method of Maximum Eccentricity \citep{Dvorak2003,Erdi2004,Vanko2013,Gajdos2019} are examples of stability indicators of this kind.  

Recently, \cite{Laskar2017} presented how AMD-stability can be used to establish a classification of multiplanetary systems. They showed the planetary systems that are long-term stable because they are AMD-stable, and those that are AMD-unstable which then require some additional dynamical studies to conclude on their stability. The authors selected the 131 multiplanetary systems (from The Extrasolar Planets Encyclopaedia database\footnote{\url{http://exoplanet.eu/}}) for AMD-stability classification. The studied sample consisted of 48 strong AMD-stable, 22 weak AMD-stable, and 61 AMD-unstable systems, including 5 hierarchical ones.  

The main motivation of the presented paper is to extend a classification of the multiplanetary systems using a different approach like in the case of \cite{Laskar2017}. We adopted the MEGNO criterion to assess the stability of exoplanetary systems. Machine-learning algorithm SPOCK \citep{Tamayo2020} was used as an additional method. Simple and relatively fast characterization of system stability shows the sample with interesting dynamics. The systems which we marked as chaotic, could be very attractive for further detailed studies in the terms of other dynamical phenomena (e.g. strong orbital resonances). Moreover, the presented comparison of results obtained using three different techniques (AMD, MEGNO and SPOCK) could certainly be beneficial in a selection of the used method in other stability studies.

The paper is organized as follows. Section~\ref{stab} provides a brief description of used indicators -- MEGNO and SPOCK. In Section~\ref{sim}, criteria for systems selections and numerical simulations setup are given. The results are described in Section~\ref{res} and concluded in Section~\ref{conclus}.

\section{Stability and chaos criteria} 
\label{stab}

Distinguishing chaoticity from regularity is the key task in the study of dynamical systems in general. The presence of chaos limits the ability of the accurate prediction of the system evolution on a long time scale. For planetary systems, chaos could sometimes mean also possible instabilities and system disintegration. This could also be used for evaluating the quality of planetary parameters set obtained during the fitting of observational data (e.g. with a combination with the Monte Carlo method).

Several methods and indicators have already been developed to determine the chaotic or regular nature of the system. The usage of a particular method is influenced not only by its suitability for a studied problem but also by its implementation in some software package that is easy to use. Basic information about the chaos indicators used in this study is given in the following sections.

\subsection{Lyapunov exponent and Lyapunov time}

One of the basic concepts, how to study the chaotic behaviour of a dynamical system, is the determination of the spectrum of its Lyapunov characteristic exponents \citep{Skokos2010}. These exponents asymptotically characterise the average rate of growth of small perturbation to the solution of a studied dynamical system. For determination, if the system is chaotic or regular, the calculation of the largest of Lyapunov exponents (called the maximal Lyapunov exponent) is sufficient. The definition of the maximal Lyapunov exponent could be written as follows:
\begin{equation}
\chi_1=\lim\limits_{t \to \infty} \frac{1}{t} \ln \frac{\left\Vert \mathbf{w}(t) \right\Vert}{\left\Vert \mathbf{w}(0) \right\Vert},
\end{equation}
where the norm of deviation vector $\left\Vert \mathbf{w}(t) \right\Vert$ characterize the size of initially small perturbation at time $t$. If the studied system is chaotic, maximal Lyapunov exponent $\chi_1 > 0$ and any perturbation increases exponentially, i.e. the nearby orbits exponentially diverge. For regular one is $\chi_1 = 0$ and the separation between nearby orbits growth linearly. 

From the maximal Lyapunov exponent $\chi_1$, the Lyapunov time of studied systems could be calculated using the well-known relation:
\begin{equation}
t_L = \frac{1}{\chi_1}.
\end{equation}
The Lyapunov time gives an estimation of the time after that a system becomes chaotic and two nearby orbits start to diverge by a factor $e$. Thus, any small change of initial parameters leads to different behaviour of the system after this time. 

In real calculations, we could not simulate the system to infinite time and the obtained Lyapunov exponent is only the approximation of the exact exponent $\chi_1$. Therefore, the non-zero Lyapunov exponent does not indicate chaotic behaviour. For that reason, the value of the calculated Lyapunov exponent is not sufficient to determine the dynamic nature of the studied system. However, it can be used to construct maps of stability showing which configuration is more chaotic and which is a more regular one \citep[e.g.][]{Payne2013}. The calculated Lyapunov time from this simulation is only the lower estimation in the case that the system does not already indicate any signs of chaos or instability (e.g. its disintegration).

The calculation of the Lyapunov exponent performed using variational equations is implemented in the package \textsc{Rebound} \citep{Rein2016} which we used for that purpose.

\subsection{Mean Exponential Growth factor of Nearby Orbits -- MEGNO}
\label{megno}

Indicator Mean Exponential Growth factor of Nearby Orbits (MEGNO) was designed to distinguish between stable and unstable orbits in a similar way as the Lyapunov exponent but in a shorter time \citep{Cincotta2000}. For general orbit $\varphi (t)$, MEGNO is defined by the equation:
\begin{equation}
Y \left( \varphi (t) \right) = \frac{2}{t} \int\limits_{0}^{t} \frac{\dot{\delta}\left( \varphi (s) \right)}{\delta\left( \varphi (s) \right)} s~{\rm d}s,
\end{equation}
where $\delta\left( \varphi (t) \right) = \left\Vert \varphi' (t) - \varphi (t)  \right\Vert$ gives the distance between two initially nearby orbits. It could be shown that the time average of $Y$ asymptotically behaves as
\begin{equation}
\label{eq:megno}
\left< Y \left( \varphi (t) \right) \right> \approx at+b,
\end{equation}
whereas $a=\chi_1/2$ and $b\approx0$ for chaotic orbit ($\chi_1$ is maximal Lyapunov exponent) and $a=0$ and $b\approx2$ for stable quasiperiodic orbit \citep{Maffione2013}. Value of $b<2$ could indicate stable periodic orbit.

In practical finite-time simulations, we mostly obtain only the value of MEGNO $\left< Y\right>$ at the end of the simulation. The significant advantage of MEGNO is that its value (with comparison to 2) is directly an estimation of the chaoticity of the studied system. Moreover, a linear fit of MEGNO values on some time interval at the end of the simulation could be used to calculate the Lyapunov exponent which is twice of obtained linear coefficient as can be seen from the previous asymptotic approximation.

However, we should stress that this estimation might not exactly implicate how the system will behave on the infinite time scale. The system which looks to be stable during finite-time simulation could always start to be chaotic on the longer time scales. Once the Lyapunov time of the stable one is estimated, we could say that the system is stable at least during this time. Therefore, a longer simulation gives a better estimation. We set the duration of our simulation long enough to detect most of the possible chaotic exhibitions. 

Used package \textsc{Rebound} calculates the MEGNO indicator together with the Lyapunov exponent in one simulation which shortens the necessary computing time for all simulations performed by us.

\subsection{SPOCK}
\label{spock}

Additionally, we used a new technique to classify the stability of planetary systems called SPOCK \citep[Stability of Planetary Orbital Configuration Klassifier;][]{Tamayo2020}. This method uses a machine-learning model based on gradient-boosted decision tree algorithm XGBoost \citep{Chen2016}. Using the very short integration, only $10^4$ orbits of the inner-most planet, the SPOCK can predict the stability of the system on the long-timescale ($10^9$ orbits). Because of the short numerical simulation, this method is very fast in comparison
to classical methods (such as the Lyapunov exponent and MEGNO mentioned above). However, we obtain only the probability that the system is stable on this time scale.  

The SPOCK method is designed only for systems with three and more planets. Unfortunately, the majority of our studied sample (see Sec.~\ref{sim}) consists of two planets. We could use this method only for 49 systems. A detailed comparison between SPOCK and classic chaos indicators is given in Sec.~\ref{res-spock}.

\section{Numerical simulations}
\label{sim}

\subsection{Systems selection}

Almost 850 multiplanetary systems are known to the present day (end of August 2022). However, the parameters of only a fraction of them are already determined. This is mainly because most of these systems were discovered using the transit method by the \textit{Kepler} mission. Their parent stars are too faint for spectroscopic observations in general. For this reason, independent discovery confirmation and planetary mass determination using the radial velocity method are impossible. If no TTV signal is detected in the system, we could not get any information about the masses of planets. Because the mass is a key initial parameter for any N-body simulation, we had to rule out all of these systems without planetary masses. The number of possible systems was reduced to about 40\%. 

An additional important input parameter is the planet location along its orbit at the simulation start to make the calculation more realistic. The correct mutual locations of the planets could be calculated from their times of passing some specific point of the orbits. For transiting planets, the transit midpoints could be used. If the exact shape of orbit with all of its parameters is known (e.g. planets discovered with radial velocity methods), the epoch of periastron is a good reference time. In addition, there are some systems with planets discovered by multiple methods with different, incompatible reference times which could not be used together and mixed. More than 120 systems were excluded at this step. 

Some multiplanetary systems are also a part of binary or multiple stellar systems. Additional star(s) could significantly change the orbits of planets. Moreover, simulations of such systems are not as trivial as in the case of single-star systems. This was the reason for including only single-star systems in our study. 

Finally, we had selected 180 systems that satisfied our criteria. Moreover, two systems had to be excluded during the simulations because of the very long necessary computational time (a few months). The main reason is a high period ratio between the outer-most and inner-most planet which influences the parameters of simulation (see below). One could expect that mainly two-planet systems with similarly large period ratios should be stable. The mutual interactions between the pair of very distant planets are extremely small.  

Other orbital elements could be calculated one from the other or assumed their values without a large or no effect on the results. The orbital period could be calculated from the semi-major axis and vice-versa. In many cases, transiting exoplanets do not have any information about their orbital eccentricity. However, these planets are often very close to their parent star where the effect of tides and orbit circularisation is strong. Their eccentricity could be small \citep{Xie2016,He2020} and we assumed that the eccentricities of such planets are equal to~0 (if the eccentricity is not already known). We can also expect that the mutual inclinations between planetary orbits mainly in the compact systems which are the biggest part of simulated sample are small \citep{Fang2012,Ballard2016,He2020} and could be neglected in cases with unknown orbital inclinations.

All information about studied systems and parameters of the planets were collected from NASA Exoplanet Archive\footnote{\url{https://exoplanetarchive.ipac.caltech.edu/}; based on the update on October 21, 2021; \cite{EA}.}.

\subsection{Simulation setup}

We have used package \textsc{Rebound} \citep{Rein2012} to perform numerical simulations of selected systems.  All bodies (a parent star and known planets) were considered as mass-full particles with mutual interactions. Therefore, our task was to solve the general N-body problem.

During the simulations, the Lyapunov exponent and MEGNO indicator were calculated. For that purpose, the package \textsc{Rebound} uses the method of the second-order variational equations \citep{Rein2016}. 

In order to obtain reliable results of chaotic behaviour of the system together with good Lyapunov time estimation, the long-term simulation with fine time step is needed. We used orbital periods of planets to set the simulation parameters uniformly. This setup allows us to compare the stability of studied systems with each other. The initial simulation time step was set to a thousandth of the orbital period of the inner-most planet. This time step corresponds to the change of mean anomaly of this planet by a value of 0.36 degree (for other planets is the value, of course, even smaller) which guarantees high precisions of the simulation and minimizes the possible computational errors resulting from large time steps. The maximum simulation time was set to 10 million orbits of the outer-most planet (i.e. to $10^7$ of its orbital periods). This time is sufficiently long for the possible chaotic behaviour of the system to occur. We ran various shorter or longer simulations to detect chaotic features and tested the necessary simulation time. In many cases, the simulations covering only $10^4$ or $10^5$ orbits gave acceptable results. Moreover, the dynamics of rapidly chaotic systems is the more interesting than stable ones or these which are chaotic on very long time scales (more than simulation time). If the system started to be chaotic and the value of MEGNO exceeded 10, the simulation was terminated before reaching the set maximum time to speed up our calculations. The value of MEGNO at the end of the simulation was directly used to classify if the system is chaotic or not.

We performed only one simulation per system. In this way, we assume that the parameters of all exoplanets are correctly determined without any uncertainties. This could be a very strong assumption in some systems. Generally, performing multiple simulations for each system which sample distribution of planetary elements from fitting observational data (e.g. using the Monte Carlo method) could provide a better overview of the system stability. On the other hand, a similar study would be a few times more time extensive and therefore not very reasonable. One could perform this detailed stability analysis for some individual systems. From our point of view, only systems labelled as chaotic (details in next section) should be interesting for detailed study which would bring some valuable results. However, it is beyond the scope of this manuscript where our main motivation was only to detect such systems. 

Separately, the SPOCK stability probabilities were calculated for each system. The only input for this calculation was the parameters of the planets and the parent star.

\section{Chaoticity and stability of studied systems}
\label{res}

From the studied sample of 178 multiplanetary systems, the big majority (74.2\%) was stable on the considered time-scale of $10^7$ planetary orbits. Only 46 systems showed chaotic behaviour. In many cases they started to be chaotic very fast -- after a thousand orbits or earlier. The obtained results (value of MEGNO, estimated Lyapunov time and SPOCK probability) for all studied systems are listed in Table~\ref{tab:res}. Note that the value of MEGNO close to 2 indicates a stable regular system and MEGNO significantly greater than 2 chaotic one (see Sec.~\ref{megno}). We set the value of MEGNO 3 as a critical value between stable and chaotic systems. This limit eliminates the possibility of false classification of the system with MEGNO slightly larger than 2 as a chaotic one. On the other hand, the real chaotic systems have the MEGNO significantly greater. As mentioned earlier (in Sec.~\ref{stab}), we should look at the stability of the system and its Lyapunov time as an estimation. We call the system stable if it is stable during the simulation (or generally on the scale of its Lyapunov time). 

\begin{table*}
\begin{center}
\caption{Chaos indicator MEGNO and Lyapunov time as a result of the simulation of individual systems and probability of stability based on SPOCK classifier. For each system, the number of planets is given. Lyapunov-time scale is expressed in years and the orbital periods of the outer-most planet. The full table is available as a supplementary material to this manuscript.}
\label{tab:res}
\begin{tabular}{cc|ccc|c}
	System & Planets & MEGNO  &    \multicolumn{2}{c|}{Lyapunov time}     & SPOCK  \\
	       &         &        &        (yr)         &      (periods)      &        \\ \hline
	24 Sex &    2    & 17.068 &  6.826$\cdot10^0$   & 2.821$\cdot10^{0}$  &  ---   \\
	47 UMa &    3    & 17.812 & 7.482$\cdot10^{5}$  & 1.950$\cdot10^{4}$  & 0.6335 \\
	AU Mic &    2    & 2.000  & 5.513$\cdot10^{10}$ & 1.067$\cdot10^{12}$ &  ---   \\
	\dots  &  \dots  & \dots  &        \dots        &        \dots        & \dots  \\
	nu Oph &    2    & 2.002  & 4.141$\cdot10^{10}$ & 4.745$\cdot10^{9}$  &  ---   \\ \hline
\end{tabular}
\end{center}
\end{table*}

\begin{figure}
\includegraphics[width=0.9\columnwidth]{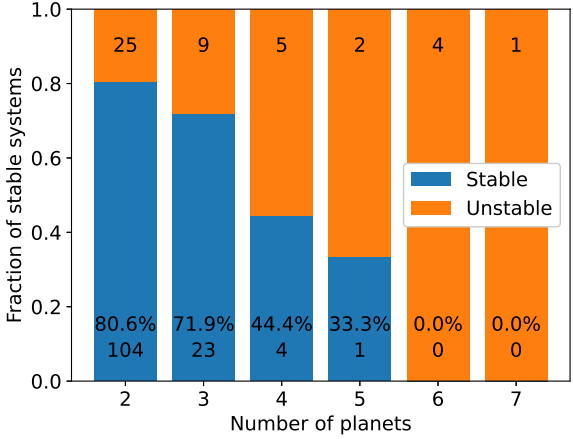}
\caption{Fraction of stable systems as a function of the number of planets. Numbers of stable and unstable systems are given on the bottom of columns or top of them, respectively.}
\label{fig:number}
\end{figure}

The number of planets strongly affects the possibility of chaotic behaviour of the whole system (see Fig.~\ref{fig:number}). Most of the stable ones are small-number-planet systems with only two or three planets. With raising the number of planets, the chance that the system is chaotic is growing. In extensive multiplanetary system, the probability of strong mutual interaction (e.g. caused by the orbital resonance), which could destabilize it, is higher. However, some of them could be hierarchical stable as was noted by \cite{Laskar2017}. In such a case, the inner and outer parts of the system could be separately regular but the system as a whole is chaotic. This is also a case of our Solar system \citep{Laskar1997}. One of the hierarchical stable systems detected by \cite{Laskar2017} is the four-planet system HD\,141399 which we found to be chaotic with Lyapunov time about 5000 orbital periods of planet HD\,141399\,e (less than 70~kyr). On the other hand, there are only a few systems with many planets in our studied sample (only 17 ones with four or more planets). Therefore, our conclusions about such systems could be insignificant from a statistical point of view and could be only some kind of selectional bias. 

\begin{figure}
\includegraphics[width=0.9\columnwidth]{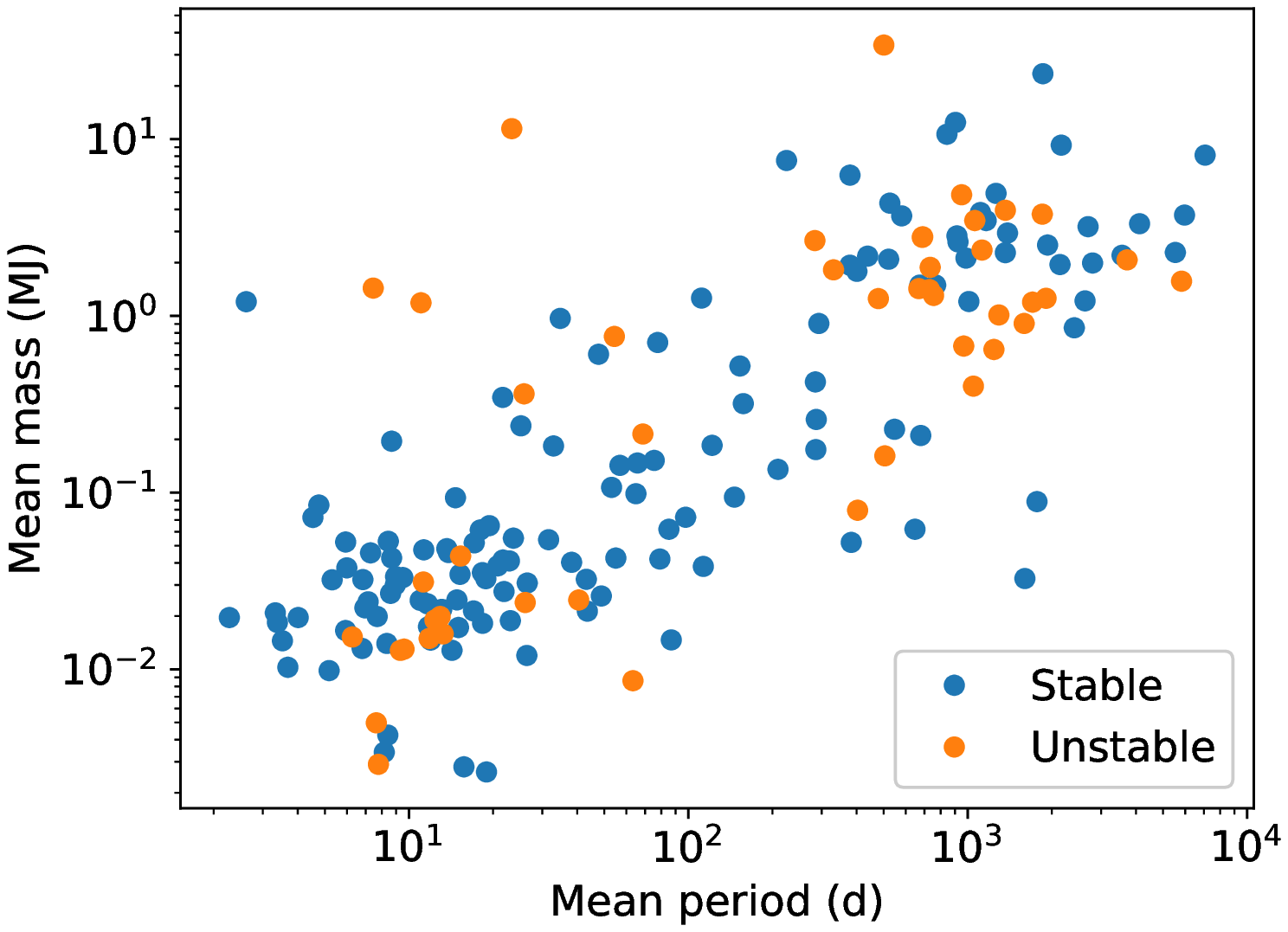}
\includegraphics[width=0.9\columnwidth]{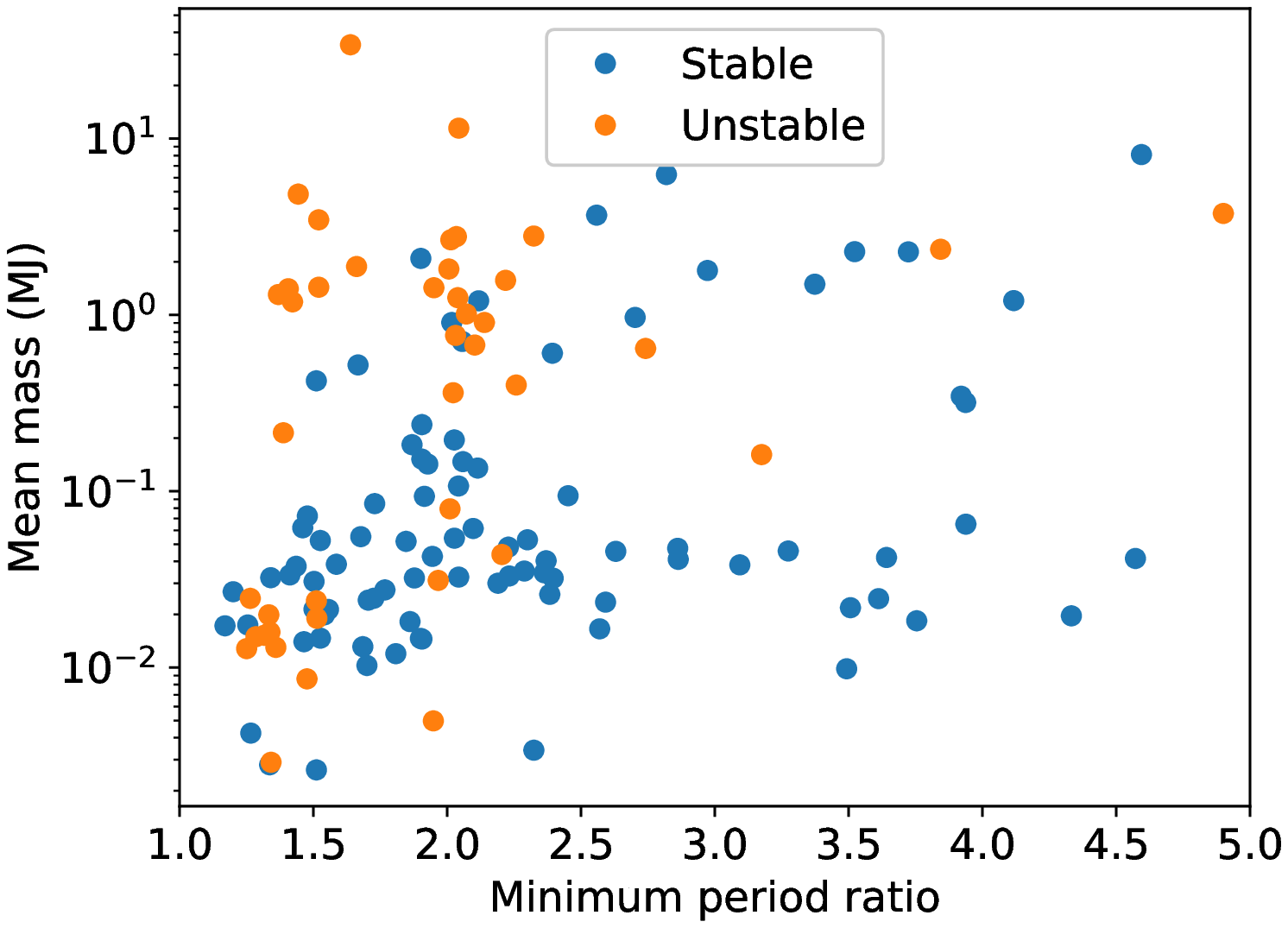}
\caption{Mass-period distribution of stable and unstable systems. \textit{Top} -- mean mass of planets in the system and mean orbital period. \textit{Bottom} -- mean mass of planets and minimum ratio of orbital periods.}
\label{fig:mass}
\end{figure}

On the mass-period diagram (Fig.~\ref{fig:mass}, \textit{top}), the regular and chaotic systems form nearly two separate groups. This could be to some extent only a result of the selectional effect or bias of exoplanets discovery methods. The mission \textit{Kepler} detected transiting planets mainly on close orbits also with Earth-like mass. The majority of long-period planets were discovered by a radial-velocity method which is more sensitive on massive planets. However, it is visible that systems with smaller and less massive planets (lower left part of \textit{top} figure and \textit{bottom} figure of Fig.~\ref{fig:mass}) are in general more stable also in a case if the system is more compact and vice-versa for more massive planets. 

\begin{figure}
\includegraphics[width=0.9\columnwidth]{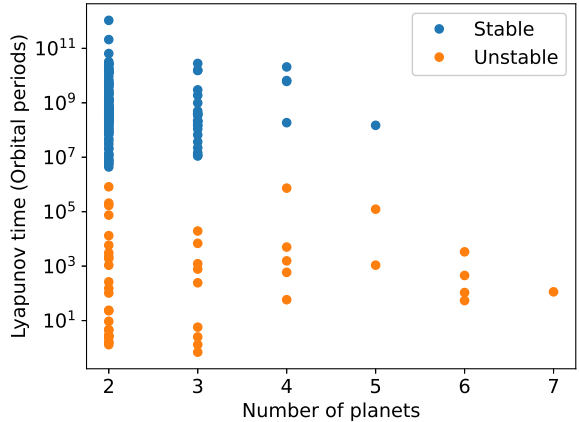}
\caption{Lyapunov time for stable and unstable systems as a function of the number of planets. Lyapunov time is expressed in the units of the orbital period of the outer-most planet.}
\label{fig:lyap}
\end{figure}

We investigated mainly the population of chaotic systems in more detail. Dynamics of some chaotic systems were already studied by different authors (e.g. GJ-876 \citep{Batygin2015}, Kepler-23 \citep{Tamayo2021} or $\gamma$\,Librae \citep{Takarada2018}). Their findings about the stability of the studied systems could be divided into two groups. Some systems are long-term stable (here, in the meaning "without collisions and disintegration") but show chaotic behaviour (mainly due to the orbital resonances between planets) -- i.e. their MEGNO is larger than 2 and Lyapunov times are short. Other ones are strongly unstable and could disintegrate (collisions and/or escape of planets) on a short time scale. This would strongly suggest that planetary parameters determined from observational data are incorrect. We ran also some additional simulations of selected unstable systems, not studied by other authors, yet, to look at the changes in the planetary orbits over a long time (Myr or Gyr). From our preliminary analysis, we could confirm the results found in the literature (mentioned above). More complex simulation and construction maps of stability which is beyond the scope of our paper would be required to determine the exact dynamics of the unstable systems.

We found the most compact pairs in the systems by calculating the minimal period ratio between all couples of planets. Figure \ref{fig:mass} (\textit{bottom}) illustrates that chaotic systems is concentrated mainly around the mean-motion resonances 3:2 and 2:1 of some pair of planets. This behaviour is observed for low-mass planets (Earth-like) and also massive ones (Jupiter-like). Due to strong system sensibility to the initial conditions close to the resonance, studying the stability of these systems in a range of orbital parameters is required to determine their real dynamic nature. The regular stable systems are distributed nearly uniformly across different period ratios. Systems with a bigger period ratio ($>5$ -- beyond the range of figure) are mostly stable.

Estimated Lyapunov time certainly separates simulated systems into two groups with a gap from $10^5$ to $10^7$ -- in full agreement with the MEGNO classification. For most stable systems, the Lyapunov time significantly exceeds the duration of the simulation. We found out some systems with the Lyapunov time close to the time limit of the simulation ($10^7$ orbits). These ones could start to be chaotic on the time scale only a bit longer than our simulations. However, on these relatively long time scales, the hypothetical chaotic behaviour could be only a result of cumulated numerical errors during the simulation. On the other hand, running some test simulations with slightly different time steps did not give significantly various results, suggesting that numerical errors have only a minimal (maybe neglected) effect. The time relation of MEGNO of a few such systems shows that the linear part of the eq. \eqref{eq:megno} start to be dominant at the end of the simulation which is related to chaos in the system. The Lyapunov time of chaotic systems approximately corresponds to the time order when the simulation was stopped (with MEGNO larger than 10; see Sec.~\ref{sim}). For some extremal chaotic cases (e.g. Kepler-23 \citep{Tamayo2021} or BD+20\,2457 \citep{Horner2014}), the Lyapunov time is only about one orbital period! The dynamics of these systems could be very interesting and required a more detailed study.

We observed a certain decrease of the maximum Lyapunov time with the number of the planets growing (Fig.~\ref{fig:lyap}). However, because we had only a few systems with many planets, a qualified decision about the real nature of this decline or only the illusion of it is not possible. 

\subsection{SPOCK probability as a stability indicator}
\label{res-spock}

We use SPOCK as an additional criterion of systems stability (details in Sec.~\ref{spock}). This indicator was applied only on systems with three or more planets -- 49 systems in the studied sample.

\begin{figure}
\includegraphics[width=0.9\columnwidth]{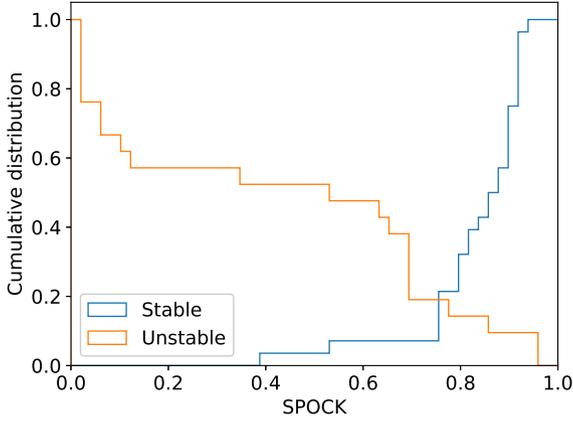}
\caption{Cumulative distribution of stable and unstable systems according to the value of SPOCK probability.}
\label{fig:spock}
\end{figure}

Figure~\ref{fig:spock} compares the stabilities of the systems determined using the MEGNO indicator with probabilities given by SPOCK. There is no complete agreement between these methods. Mainly, some chaotic systems have a relatively large stability probability (more than 50\%). If one wants to separate systems into stable and unstable ones according to SPOCK probability, the threshold should be set to around 75\% to obtain similar results as in the case of using MEGNO. On the other hand, the MEGNO stable systems have in general also high SPOCK probability (at least 80\%).

\begin{table*}
\begin{center}
\caption{Systems with different results (stability or chaos state) from analysis based on MEGNO and AMD-based study of \citet{Laskar2017}. See a detailed description of Table \ref{tab:res} and discussion in the text.}
\label{tab:diff}
\begin{tabular}{cc|ccc|cc}
	  System   & Planets & MEGNO  &   \multicolumn{2}{c|}{Lyapunov time}   & SPOCK  &   AMD type    \\
	           &         &        &       (yr)       &      (periods)      &        &               \\ \hline
	 GJ\,876   &    4    & 23.550 & 1.992$\cdot10^1$ & 5.853$\cdot10^{1}$  & 0.0000 & strong stable \\
	HD 113538  &    2    & 18.571 & 5.117$\cdot10^2$ & 1.027$\cdot10^{2}$  &  ---   & strong stable \\
	HD\,183263 &    2    & 17.552 & 2.228$\cdot10^3$ & 2.649$\cdot10^{2}$  &  ---   & strong stable \\
	HD\,187123 &    2    & 10.819 & 8.586$\cdot10^6$ & 8.225$\cdot10^{5}$  &  ---   &  weak stable  \\ \hline
	HD\,136352 &    3    & 2.368  & 6.538$\cdot10^6$ & 2.225$\cdot10^{7}$  & 0.9244 &   unstable    \\
	HD\,155358 &    2    & 2.017  & 5.723$\cdot10^8$ & 5.330$\cdot10^{8}$  &  ---   &   unstable    \\
	HD\,20003  &    2    & 2.029  & 2.787$\cdot10^7$ & 2.999$\cdot10^{8}$  &  ---   &   unstable    \\
	HD\,21693  &    2    & 2.034  & 3.346$\cdot10^7$ & 2.273$\cdot10^{8}$  &  ---   &   unstable    \\
	HD\,31527  &    3    & 2.677  & 8.957$\cdot10^6$ & 1.203$\cdot10^{7}$  & 0.8202 &   unstable    \\
	HD\,45364  &    2    & 1.998  & 3.962$\cdot10^9$ & 4.218$\cdot10^{9}$  &  ---   &   unstable    \\
	HD\,47366  &    2    & 1.988  & 4.389$\cdot10^8$ & 2.346$\cdot10^{8}$  &  ---   &   unstable    \\
	HD\,92788  &    2    & 1.945  & 4.585$\cdot10^9$ & 1.441$\cdot10^{8}$  &  ---   &   unstable    \\
	HIP\,57274 &    3    & 2.161  & 7.847$\cdot10^9$ & 6.635$\cdot10^{7}$  & 0.7695 &   unstable    \\
	Kepler-87  &    2    & 1.998  & 6.628$\cdot10^9$ & 1.265$\cdot10^{10}$ &  ---   &   unstable    \\ \hline
\end{tabular}
\end{center}
\end{table*}

We found 11 MEGNO chaotic systems (about 22\% of systems with calculated SPOCK probability) whose SPOCK predicts as stable ones. Three of them (HD\,37124, HD\,141399, Kepler-11) are also AMD unstable. We could not identify the reason for this disagreement. One hypothesis could be that the chaotic and unstable behaviour of these systems starts to play some role only on time scales longer than the short simulation time used by SPOCK ($10^4$ orbits of an inner-most planet). However, the Lyapunov time of some of them is relatively small for even detecting the chaotic nature also by SPOCK. A more detailed study which would probably compare also results obtained using different methods would be required to uncover the reason for this discrepancy and possibly the real nature of these systems.

\subsection{MEGNO vs AMD-stability}
\label{res-amd}

One of our motivations for this study was to compare MEGNO as a chaos indicator with the results of \cite{Laskar2017} using the AMD criterion. The AMD is a relatively simple analytical criterion that estimates the possibility of close encounters or collisions between the planets in the system. This method is concentrated on secular dynamics and therefore ignores the effects of mean-motion resonances which could be a source of dynamical chaos in the system. Moreover, the result of the AMD criterion should be interpreted as a worst-case scenario and the predicted collisions of planets should not occur during the real simulation. 

Our systems selection was different, and therefore only a part of them was studied by both methods. The desired comparison between MEGNO and AMD was possible for 62 systems. Only 15 of them have at least three planets and their SPOCK probability could be calculated. For many systems, we obtained results similar to \cite{Laskar2017}. In the case of 14 (listed in Tab.~\ref{tab:diff}), our results differ. Moreover, we used updated orbital elements compared to \cite{Laskar2017}. Since the direct comparison of MEGNO and AMD was not our goal, for most of the systems we have used exactly the results of \cite{Laskar2017}. For the systems with some differences in the stability prediction, we recalculated the value of AMD using the latest planetary parameters. However, we did not observe any significant discrepancy.

Three systems are AMD-stable but they start to be chaotic very quickly in our simulations. Their Lyapunov times are from about 60 to 270 orbital periods. SPOCK stability probability for four-planetary system GJ\,876 is also exactly zero which means that configuration of the system goes unstable within the $10^4$ orbits. The dynamics of this system were already studied by \cite{Batygin2015}. They marked GJ\,876 as a rapidly chaotic system with a Lyapunov time of about 50 years (approx. 147 orbits) which is 2.5-times longer than our estimation but confirms that the system is chaotic on a very short time scale. The authors together with \cite{Rivera2010} discussed that the system could be long-term stable and the chaos is exhibited by Laplace resonance between planets b, c and e. The dynamics of the other two AMD-stable and MEGNO-chaotic systems were not studied, yet. Orbital periods of two known planets in both systems look to be far from possible mean-motion resonance. Moreover, system HD\,187123 is weak AMD-stable with the possibility of collision between the inner-most planet and the parent star. The MEGNO analysis marked it as chaotic with Lyapunov time less than 10~Myr. AMD and MEGNO analyses are not entirely in disagreement, in this case.

The opposite situation is in the case of 10 systems which are denoted as AMD-unstable but MEGNO analysis claims that they are regular with Lyapunov time at least a few Myr or Gyr. The high SPOCK probability of three of them (HD\,136352, HD\,31527 and HIP\,57274) also predicts their long-term stability. The only system of these AMD-unstable ones with already studied dynamics is HD\,45364 with two planets. \cite{Correia2009} showed that 3:2 mean motion resonance between the two planets prevents close encounters and ensures the stability of this system over a few Gyr. Without this resonance, the planetary orbits, which almost cross, would be strongly unstable. \cite{Laskar2017} using AMD criterion predicted the possible collision between planets as a worst-case scenario but the observed orbital resonance never allows planets to get closer than 0.371~au \citep{Correia2009}. Even for other these AMD-unstable systems, the AMD criterion predicts the possibility of collision between planets or the parent star and inner-most planet if the eccentricities of these planets increase. The collision with a star is a case of the systems HD\,136352 and HD\,92788.
Some ones are close to resonance which could protect them from any collision (similarly to HD\,45364). Although, some of these systems could start to be chaotic and unstable on longer scales. On the other hand, the estimated Lyapunov times of most of them are at least a few Gyr which is near to the lifetime of the parent stars and the planetary systems. Therefore, it makes no sense to study if the system is stable or not on the longer time scales.

\section{Conclusion}
\label{conclus}

We used indicator MEGNO to study basic dynamics and stability of 178 multiplanetary systems from the NASA Exoplanet Archive. The studied sample has been separated according to the value of MEGNO into groups of 132 stable regular systems and 46 chaotic ones. Most of the chaotic systems are concentrated around the mean-motion resonances between planets. More planets in the system generally decrease the probability that the system is long-term stable. However, such systems could be hierarchical, i.e. could be split into two separately stable parts \citep[as noted by][]{Laskar2017}. We found out that the mean Lyapunov time of stable systems is about $10^9$ orbits, which is 100-times longer than the duration of our simulation.

The machine-learning algorithm SPOCK was used as an additional approach to predict the systems stability with three or more planets (49 studied systems). Despite some small disagreements between SPOCK and MEGNO, this technique could be very useful as the first, very fast method to determine the stability of the studied system. Mainly for very chaotic or conversely extremely stable ones, it gives excellent results. From our point of view, its main limitation is unusability for two-planet systems which compose the biggest part of known exoplanetary systems.

Finally, we compare our results with a previous similar study of \cite{Laskar2017} using the AMD criterion. From 62 systems studied in both papers, we found 14 for which results coming from MEGNO and AMD analysis are different. The main reason for this discrepancy is the absence of resonant interaction between planets in the AMD calculation. As \cite{Laskar2017} mentioned, we could not say that an AMD-unstable system is necessarily unstable without any further dynamical analysis.

One of the goals of the presented paper was to identify systems with possibly interesting dynamics for the following studies. Especially, the chaotic ones require more detailed dynamical analysis for distinguishing their real nature and all mutual planetary interactions. Excluding the work of \cite{Laskar2017}, the dynamics of only about 30 systems from our studied sample were studied in any detail in the literature. Most of the authors are concentrated mainly on some particular topics of the selected one -- stability of habitable zone \citep[][AU\,Mic]{Kane2022}, resonant interactions (e.g. \citealp{Hamann2019}, K2-146; \citealp{Panichi2019}, KOI-1599 or \citealp{Leleu2021}, TOI-178), long-term evolution (e.g. \citealp{Trifonov2020}, GJ\,1148 or \citealp{Zhang2021}, Kepler-129), system formation (e.g. \citealp{Xiu2021}, HD\,106315 or \citealp{Petigura2018}, K2-24) or construction of the maps of stability (e.g. \citealp{Wittenmyer2017}, HD\,30177 or \citealp{Baluev2014}, HD\,82943). Dynamics of only a few systems were studied detailly in the general context (e.g. \citealp{Gozdziewski2007}, HD\,160691 or \citealp{Teyssandier2022}, TRAPPIST-1). With fast and powerful computing technology will become more available, we can expect a rapidly rising number of dynamically studied systems in future studies.

\section*{Acknowledgements}
This research has made use of the NASA Exoplanet Archive, which is operated by the California Institute of Technology, under contract with the National Aeronautics and Space Administration under the Exoplanet Exploration Program.

This work was supported by the Slovak Research and Development Agency under contract No.~APVV-20-0148. 
This work was also supported by the VEGA grant of the Slovak Academy of Sciences No.~2/0031/22.
The research of PG was supported by internal grant VVGS-PF-2021-2087 of the Faculty of Science, P. J. \v{S}af\'{a}rik University in Ko\v{s}ice. 

\section*{Data Availability}
All information about studied systems and parameters of the planets were collected from NASA Exoplanet Archive \citep{EA}. Results of our stability study (values of MEGNO, Lyapunov time and SPOCK) for all analysed systems are available in the article and in its online supplementary material.

\bibliographystyle{mnras}
\bibliography{bibfile}

\section*{Supporting information}
Additional Supporting Information may be found in the on-line version of this article:\\

\noindent Table \ref{tab:res}. Chaos indicator MEGNO and Lyapunov time as a result of the simulation of individual systems and probability of stability based on SPOCK classifier. For each system, the number of planets is given. Lyapunov-time scale is expressed in years and the orbital periods of the outer-most planet.\\

\noindent Please note: Oxford University Press is not responsible for the content or functionality of any supporting 
materials supplied by the authors. Any queries (other than missing material) should be directed to the corresponding 
author for the paper.

\bsp	
\label{lastpage}
\end{document}